# Why Teach Quantum In Your Own Time:
# The Values of Grassroots Organizations Involved in Quantum Technologies Education and Outreach.

Ulrike Genenz
*RWTH Aachen University*
Aachen, Germany
ulrike.genenz@rwth-aachen.de

Neelanjana Anne
*University of California*
Irvine, California, USA

Zeynep Kılıç
METU *Middle East Technical University*
Ankara, Turkey

Daniel Mathews
*B.M.S. Institute of Technology and Management*
Bengaluru, India

Oya Ok
*Terakki Foundation Schools*
Istanbul, Turkey

Adrian Schmidt
*Institute for Technology Assessment and Systems Analysis (ITAS)*
Karlsruhe, Germany
adrian.schmidt2@kit.edu

Zeki Can Seskir
*Institute for Technology Assessment and Systems Analysis (ITAS)*
Karlsruhe, Germany
zeki.seskir@kit.edu

*Abstract*—This paper examines the intersection of goals and values within grassroots organizations operating in the realm of quantum technologies (QT) education. It delineates a fundamental distinction between the objective to provide education and the drive to democratize learning through principles of inclusivity, accessibility, and diversity. The analysis reveals how these organizations navigate their nascent stages, grappling with the dual challenge of adhering to their foundational values while aspiring for sustainable growth and development in the highly specialized field of QT. The study uncovers the strategic approaches adopted by these entities, including efforts to create educational ecosystems and foster community engagement. The research underscores the potential vulnerabilities of these grassroots organizations, particularly in relation to the longevity and evolution of their initiatives as members transition into professional roles within the quantum sector. Through this investigation, the paper contributes to a nuanced understanding of how emerging educational organizations in the QT field balance their ideological commitments with practical growth considerations, highlighting the critical factors that influence their trajectory and impact.

INTRODUCTION AND BACKGROUND

Quantum education is one critical pillar, enabling societies to make the most use of second-generation QT, by training future workforce and equipping the society to comprehend and trust in this emerging technology. Quantum education is on the one hand about providing educational programs at high-level institutions, designing curricula, and training future physicists, engineers, and quantum information scientists about exploiting quantum phenomena. On the other hand, it is about creating a broad understanding of the capabilities of QT, enabling discourses, and open innovation besides societal engagement and allowing for participation, gaining a broad quantum readiness of the future workforce and in society in general [1, 2]. QT is expected to impact society at large. Even though it does not seem clear yet, in which depth and significance this impact will be, it has been becoming apparent that QT has the potential to change the current status quo of information security and internet communication and even further there is potential for the disruption of well-established concepts as we know them [3, 4]. Over the last couple of years there has been an increasing call from researchers towards responsible innovation, participation, and engagement of the public sphere, so that the discourse of the chances and impacts of the emerging technology will be open, diverse, and enriched by various perspectives [1, 4–6]. The manifesto by Coenen et al. argues for QT to be comprehensible to different groups of stakeholders, specific, open, accessible, and responsible, considering also unforeseen or unintended consequences [2].

On the other side, there is an ongoing quantum race, where companies and countries vie for technological and competitive advantage, having national strategies and programs in place. There has been a steep increase of newly founded start-ups, patent applications and high investments in QT in recent years [7–9], contributing to a culture of competition and rivalry about intellectual property rights and trade secrets [4]. There are an increasing number of news about export restrictions of essential parts of quantum computers for reasons of national security [10–12]. According to Kop et al. this rivalry can be a driver for innovation but may be a hindrance to ethical development, as ethical, legal or social implications (ELSI) are not incentivized [4]. Additionally, this trend has a strong effect on the disparity, inequitable access to QT and essential resources, it worsens the so-called quantum divide [13], a concept that refers to a significant gap in access and utilization of QT causing inequality. A study by Gercek and Seskir provides a detailed image of a quantum divide in science, in technology, between countries, and between societies [14]. Due to that mentality of 'the winner takes it all' there are different approaches to creating QT in different regions and some countries have nor the resources nor access at all to existing technological approaches, tools, and knowledge, increasing a gap between the 'haves' and 'have nots' even further [13]. That different perspectives could be a great game changer for the development of QT is an ongoing discussion in literature, with the ideas of knowledge being exchanged more freely between countries of similar values and synergies of different approaches, methods and tools being used effectively. Scientists are talking about bridging the quantum divide [3, 4, 13], emphasizing the advantages of participation and engagement of diverse perspectives in the innovation process to fully understand, utilize, and benefit from QT [4, 13, 14]. Some argue that a diverse and inclusive responsible innovation of QT facilitates an effective and sustainable development, actually developing and using the technology for real-world purposes on a larger scale [4]. Additionally, democratic engagement would create broad access to this emerging technology, building up a comprehensive knowledge base, reducing hesitancy and increase awareness, which would support entire ecosystems and market formation efforts [5]. Another benefit, scientists argue for, is that societies would be a lot stronger in facing risks of dual or malicious uses of quantum applications, that could harm societies on another level risking a destabilization and undermining trust in institutions [15].

The inclusion of quantum grassroots organizations, that are a form of upstream engagement of the public and are an integral part of civil society organizations [5, 16], into national quantum strategies and policy makings finds so far only little attention [16]. But, there is a strong focus on building up educational programs. Scientists, the industry and governments are addressing a shortage of quantum literate talents, pushing to build human capital in quantum. Over the last ten years, many education initiatives have evolved in the USA, Canada, Europe, India, Singapore, Japan, South Korea, Australia, South Africa, and many more [17]. These educational programs focus mainly on quantum physics, quantum engineering, and quantum information science targeting STEM students and are mostly run by well-known universities. But studies show that they lack a gender balance and an inclusion of people of color and minorities, as well as low-income students [1, 18–20]. Despite a well-known workforce shortage and calls from ethicists about a need for diversity and inclusion in the quantum sphere, the progress seems slow. According to Wolbring, academic literature concerned about QT rarely to not at all addresses topics that relate to social implications as well as equity, diversity, and, inclusion [21].

Kaur et al. have analyzed different pathways that facilitate quantum education, showing that next to academic degree programs, there are also shorter trainings that are fitted to the demands of the job and real-world projects [17]. Further initiatives with a focus on diversity and inclusion are established, for example, online degrees, to be accessible to more students. The European Quantum Flagship program, a research and innovation initiative by the European Commission, has started DigiQ, a virtual education program in addition to any master's degree [22]. Furthermore, the Quantum Flagship is monitoring the progress of QT and has dedicated one scorecard entirely to education, evaluating outreach, the offered training modules, adoption as well as diversity and equity. The numbers reflect a significant increase in educational efforts over the past two years, with ambitious goals set for 2030 [23]. While academic programs are crucial for building a quantum-ready workforce and are in high demand for quantum industry job profiles [24], it's important to recognize the need for more diverse efforts. These include attracting quantum enthusiasts, offering learning programs beyond university through experience, learning quantum skills through training courses, and contributing to quantum communities [17].

RESEARCH QUESTION

Quantum grassroots organizations create global and local quantum communities, contribute to quantum education, and gain more quantum enthusiasts. Over the last decade the organizations have become part of the quantum ecosystem being in the position to support a quantum literate workforce and science communication to a larger audience. This study 'Why teach quantum at your own time' raises the question of why people voluntarily contribute to quantum education: Do they align with current debates regarding the workforce gap, quantum divide, and calls for responsible innovation? Are they pushing towards societal change by popularizing quantum education or are they rather eager learners supporting each other? With the study, we want to capture their reflection on their goals (what do they aim for?), values (what motivates them?) and teaching and learning methods employed (how do they do it?), providing a comprehensive understanding of their contribution to quantum education and quantum ecosystems.

In this regard, we employed a survey and conducted semi-structured interviews (more on these in the Methodology section). Our main questions in the survey and interviews were:

(1) "What is the main goal of your organization? Could you please provide us with some more information about the intention of the organization?"

(2) "What are the most important values of your organization? Please list at most 5 values your organization works with. Please order the values you just named from highest to lowest."

(3) "What are the teaching/learning methods that your organization employs?"

In the following section, we introduce the methodology of the mixed-methods approach, providing an overview of the study's concluded design, how the data was evaluated, and showing the limitation to the study. Afterwards, the results section mirrors our three main questions and is enriched by findings in the organizational set-up and demographics. Moreover, comparing goals and values between regions, we found some trends of local characteristics, which are described at the end of the results. Finally, there is a discussion section that aligns the results with current findings in research and is followed by a conclusion of the study.

METHODOLOGY

**Research Design**

Through the use of mixed methods, this study employed a survey and interviews with quantum grassroots organizations. The selection of these organizations was carefully conducted through extensive searches on LinkedIn and Google building upon previous projects, one on mapping the Communities of QT [20] and another one that ran as a summer project in 2023 with volunteering interns, called QInterns, under the QWorld's 2023 summer internship program. A list was generated of existing grassroots organizations to which the survey was distributed. The survey included 13 questions in total that addressed demographics, personal membership, motivation, and organizational goals in the first section, whereas the second section contained the organization's perceived five core values, which participants were asked to rank, teaching and learning methods, and outlook. The interviews were primarily with the organizations' founders and established members, and the questions were congruent with the survey, aiming at a better understanding of the survey responses. The interviews were conducted in two rounds. The first round of interviews was aimed at organizations that were globally active and well-established as grassroots organizations. The second round of interviews focused on the geographical location of grassroots organizations to gain a better understanding of regional differences. Three organizations were identified as focus groups and observed in more detail, aiming at a better understanding of congruity of their named values. They were selected based on their global set-up.

**Data Collection**

The survey, which was designed using Qualtrics software, was distributed to 66 grassroots organizations around the globe, to which 18 organizations participated via the survey, of which only 15 responded to all questions. There were 20 interviews with participants from 17 organizations chosen out of the 66 due to their size, location, and availability for interviews during the interview phase in November 2023. From the two big players, QWorld, with their 32 local spin-offs called QCousins, and OneQuantum, with 12[1] communities worldwide, there have been three local organizations of both chosen. The remaining 11 organizations in the interview set act either purely global, without local spin-offs or only locally, focusing on their region or state. Out of all the organizations interviewed, three global organizations were interviewed twice, addressing two different long-term members or a member and a founder. Overall, there were 22 organizations participating by survey and interview. The grassroots organizations were from Africa (6), Asia (5), Europe (3), North America (7), and South America (1). Participants were invited to partake in a 60-minute interview via Zoom, and each session was recorded and transcribed using AI f4x transcription software.

**Evaluation**

The survey data was analyzed using Excel and SPSS, while the transcripts were carefully evaluated with the assistance of MAXQDA. The interview responses were analyzed with codes based on the participants' statements. To further refine the responses, sub-codes were created from the main codes. These categories were then linked to the survey responses to provide a comprehensive and detailed analysis. Standardizing the spelling of all terms ensured consistency throughout the analysis.

*Demographics*: The data on organizational structure and demographics was evaluated from the survey and the interviews and also compared between the statements of different members of an organization, where available. Where statements diverged, the mean value was calculated. The questionnaire did not contain a specific question on the year of foundation of the organizations, but the data was available in 12 interviews. Missing information on the founding year was researched on the Internet via the organizations' websites and LinkedIn, whereby the founding year was researched for all 66 grassroots organizations and presented as a supplement. The list of all grassroots organizations was extended again in 2024 and 3 additional grassroots organizations were found. Of the 69 organizations, the year of foundation could not be determined for 3 organizations.

---

[1] According to the OneQuantum website, there are 22 chapters, but only 12 could be identified during the study.

*Goals*: The objectives mentioned were summarized according to the most frequent determining factors of all mentions. Since grassroots organizations mentioned between 3 and 5 goals in free text, it was possible to assign multiple mentions to an organization. In the survey, only 12 out of 15 organizations responded to the question about goals, so together with the interviews, a total of 20 organizations responded. First, the goals in the survey data were summarized according to the most frequently cited determinants. Then, the interview data on goals were assigned to codes already derived from the survey mentions. Codes were expanded from all interview data. Goal scores from both data sources were compared and adjusted to achieve standardization. Once the final scores were established, the two data sources were compared again. A data set for quantum education allowed for a more detailed breakdown of different educational goals due to the data available. This is shown in Figure 3 in the results section.

*Values*: The data on values was counted by frequency in both the survey and interviews but also evaluated by ranking, showing the frequency count of the highest value. Some survey participants mentioned value pairs like 'openness & accessibility' which was counted as two separate entries, counting 28 entries from 22 organizations.

*Teaching and learning methods*: The data on teaching and learning methods were summarized based on the most frequently mentioned factors by participants in both the survey and interviews.

RESULTS

**Organisational Structure and Demographics**

Data on size was available from 21 organizations. 25% of them (5) have less than 10 people, and almost 60% (12) have a core of 10 to 20 people. Less than 20% (4) have more than 20 people. A vast majority of the people working within these entities are volunteers participating in university or school and define its stability as they transition into various parts of their career.

The interviews provided insight into their organizational structure: There are global organizations that work and act globally and/or set up local divisions, connecting through digital platforms like Slack or Discord. On the other side, there are local organizations that act in their country or a region of the country and do not belong to a global network. Most of the organizations interviewed showed signs of professionalizing their work structure with different departments and working groups, with strategic planning, task orientation, and regular meetings. The following statements show their professionalizing efforts:

*"In our team we do quarterly meetings where we […] think about, what are our goals for this next three months? What can we measure those by? […] the way that our meetings typically go is: I [founder of the organization] will start by giving my overview on how the quarter has gone. I'll celebrate our successes. I'll talk about some things we could have done better and just talk about our general mission and reminding people of what that is, and then I'll open it up to everyone. We have like a whiteboard sticky note session where everyone can talk about what they think is really important to get to this quarter."* (Participant 13)

*"Those 30 people that work on the content, they're divided into five teams. […] we have people specifically for planning, webinars and recording interviews. We have people specifically for creating the marketing content that we use on Instagram, high level management and organization of that. Also, directing educational program development by doing internal and external outreach."* (Participant 3)

But not all organizations were in such thriving conditions. Grassroots organizations depend mainly on volunteers, facing a steady in- and outflow of members when degrees are finished and careers are pursued, leading to transitions of founders and key figures in the organization.

Participant 20 reported that the organization *"is in hibernation at the moment, so we have not organized any collective events for more than a year. It's mainly because me and my co-founder have moved, so we had our own problems moving to other countries and settling down"*.

Overall, the interviews showed little evidence of transition structures for members in key positions.

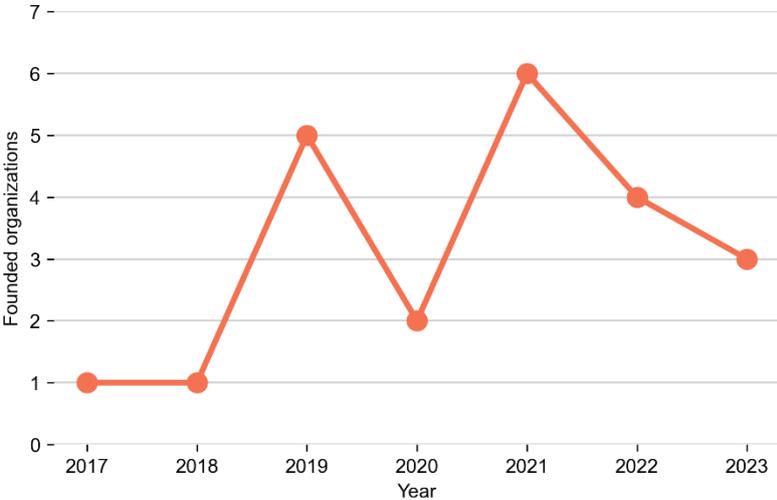

*Figure 1: Number of founded grassroots organizations per year (participants of the study).*

Looking at the age of grassroots organizations, they are mainly quite young. Most of the organizations were founded in and after 2019 and 2020, which can be explained by the creation of global parent organizations around 2019 and the creation their local spin-offs around the world. The two largest global organizations, QWorld and OneQuantum, both founded in 2019, have since created about 44 local spin-offs. Both graphs, Figure 1 and Figure 2, show a very similar curve pattern, comparing the number of study participants and all identified grassroots organizations, with a peak in 2021.

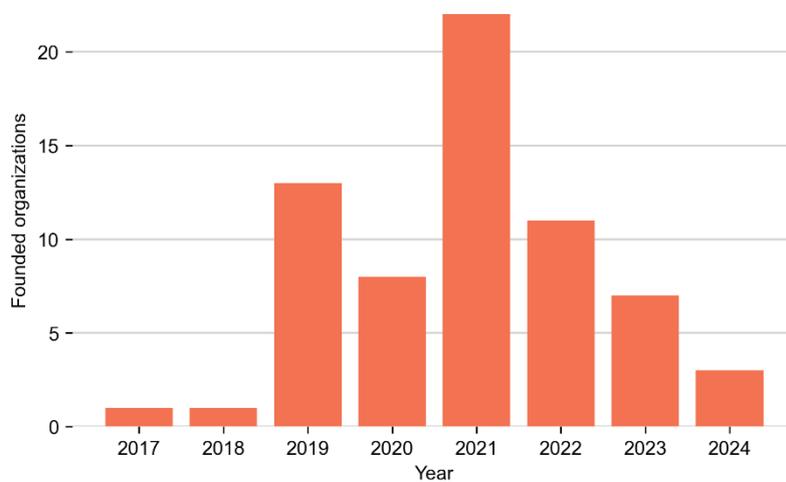

*Figure 2: Number of grassroots organizations founded per year from the full list of 69 grassroots organizations, minus 3 organizations without data for the year of founding.*

The growth of new organizations coincides with several factors. First of all, some organizations report that they became interested alongside with the hype around quantum computing, for example, when Google announced that it had achieved quantum advantage in 2019. [25] Another prominent attraction factor seemed to be the global pandemic of COVID-19, which pushed education massively into online spaces, and personal periods of isolation and/or quarantine led to an increased use of online communication and social networks [26]. IBM for example offered their Qiskit Global Summer School in a virtual open-access space, reaching over 4,000 participants [27]. The following statement indicates that the IBM initiative was a critical factor in the creation of the organization.

*"We, I and a bunch of other students learning quantum or involved in Quantum met at the first global summer school in 2020. We all met at the Discord server that they had and eventually we all got to talking about creating communities and creating spaces for learning more about quantum effectively and in a way that is open source and inclusive and eventually so on and so forth. We just created the space with all these resources online. And it's not just us, but we also invite everyone in the community and everyone online to participate as well and come in and join and contribute to resources and contribute to learning together."* (Participant 6)

**Goals**

In the field of quantum science and technology, grassroots organizations have emerged as influential facilitators in education as well as creating opportunities for learners, both being the most frequent mentioned goals. Figure 2 shows the full distribution of mentioned goals.

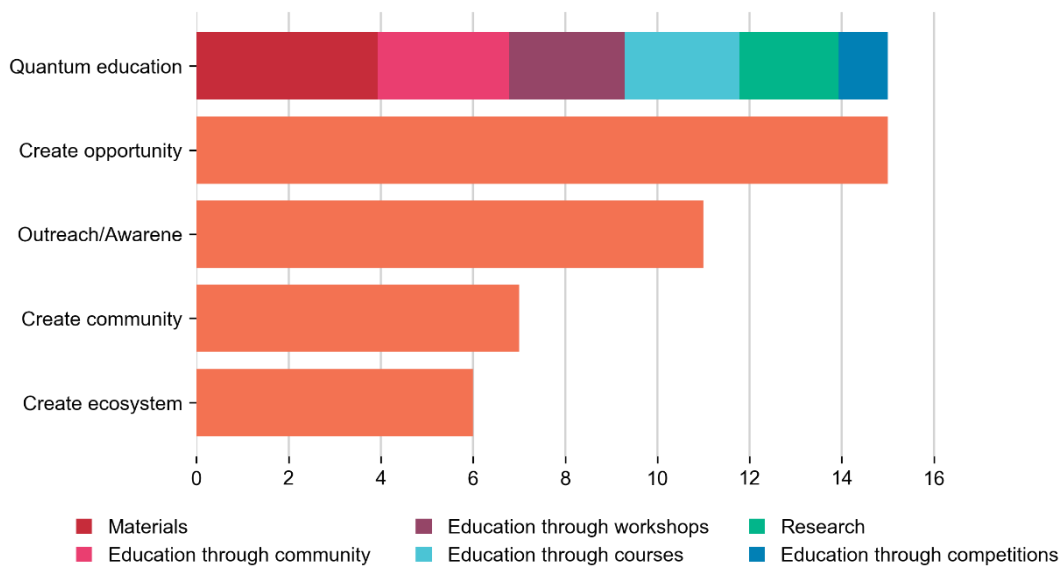

*Figure 3: Goals of grassroots organizations (N=20), sorted after frequency and grouped according to similarities. The goal quantum education was broken down into sub-goals due to data availability, showing their focus on different educational styles.*

Our research has found that grassroots organizations demonstrate a strong commitment to **quantum education** and to the development and provision of accessible and comprehensive educational materials, supporting learners across a range of educational stages. The organizations offer diverse educational possibilities, addressing different learning styles: theoretical to practical, social to independent, and competitive to collaborative learning, which will be addressed further down, when talking about teaching and learning methods. Based on the different methods in education the graph on quantum education was already split, showing the different results of educational styles. It shows that the distribution of mentions is quite similar, offering a diverse range of learning about quantum. Most of the grassroots organizations (n=15) are dedicated to quantum education and aim to reach STEM and non-STEM students at all levels, academics, professionals, and the general public. They aim to support the quantum journey for all learners, guiding them along the way and nurturing talent from an early age.

Another goal mentioned as often as education was **creating opportunities**, which is an aggregation for mentoring and supporting students, providing ample opportunities to learn, and building a successful career in quantum. By emphasizing this goal, grassroots address various needs in quantum. By creating access to resources and supporting students in their learning, they contribute to democratize learning. Additionally, they aim at fostering talent from different backgrounds, addressing diversity gaps in quantum education and aiming at people to learn about quantum to eventually contribute to a to a skilled workforce and create impact. Over half of the surveyed organizations are concerned about the lack of diversity in the quantum workforce. Studies reveal that STEM graduates from educationally marginalized groups are underrepresented in terms of race, ethnicity, sexual orientation, and gender [18, 20]. A study from 2024, that is mainly concerned with quantum information science (QIS) in the USA, showed that QIS education programs are less available in rural states and at institutions that serve more low-income population in the USA [19]. Grassroots efforts aim to create opportunities for minorities such as women, people of color, and other minor backgrounds and actively placing them in niche areas of science such as QT. They create a supportive and safe environment and target underrepresented people. The inclusivity objective is an important contribution to democratizing access to QT [5, 19, 28], not only in terms of fair educational opportunities but also providing chances for societies that are less technologically advanced and are not as equipped with the resources to take advantage of the potential benefits that QT can offer [4, 13, 29]. Organizations from disadvantaged regions and the Global South address inclusivity as part of global quantum development at a systemic governance level, meaning that these organizations aim at creating opportunities for their society on a global scale.

> *"IBM experience is not accessible in some African countries. […] We don't have a quantum computer in Africa yet. So, how can we mitigate this to make sure that at least the interested people have the opportunity to learn? We try to create equality, whereby, we don't have the same resources as everyone else, but we do want to learn. We do want to also be part of this quantum revolution."* (Participant 9)

**Outreach and awareness** raising are inclining to attract members, media attention, and funding. Over half of interviewed and surveyed organizations named outreach as a goal to gain visibility, opportunities in partnerships and sponsoring and "*raise awareness about this emerging field of technology*" (Participant 17) within society as one participant commented or to "*increase the number of students who are engaging with this exciting field*" (Participant 15) like another participant responded. There was no particular group of grassroots organizations that declared outreach as a goal, but it can be said, that all grassroots organizations are concerned with outreach and awareness, even if they do not state it as their main goal.

Some organizations aim for a strong quantum **community** as a place to support each other on the individual quantum journey ("learning together"), to collaborate and connect with like-minded people, and to inspire members. Organizations that focus primarily on building a quantum community organize goals such as quantum education and opportunity creation in different ways. Learning becomes more of a social event, with less material and more focus on community facilitation, getting to know each person's context, and building community exchange. Because of this setup, they provide a platform for new members to join and contribute content, as well as get questions answered by the crowd. Their communities also strive to create a diverse, safe, and inclusive network.

> [We want] *"to support underrepresented groups, to keep them in quantum and to give them a network of people like them."* (Participant 18)

The goal **creating an ecosystem**, which was highlighted by six organizations, was often but not exclusively used by organizations from geographical regions that do not yet have a strong presence on QT. Some organizations from the Global South aim to create an ecosystem and make an impact even before providing quantum education to youth or building a community. Being faced with the challenge that key stakeholders such as academic and governmental institutions and the business community are not fully aware of the possibilities of QT, the goal of grassroots organizations shifts towards governmental, academic, business, and societal awareness.

Talking with founders of grassroots organizations (n=7) about their motivation shows the fine line between the goals of the organization and the values behind the organization. The founders had their own reasons for starting an organization in quantum education. One reported, that there was a lack of structured content for early beginners, another founder did focus on creating a community for learning together. One founder from Africa said, "*If you want to master something, you need to teach it*." (Participant 17), showing that this motivation was based on personal intend to deepen the understanding of the field and take on some form of leadership. Whereas these motivations were rather personal, another founder replied on a more structural level by explaining that the rapid pace of technological development in quantum is not being adequately addressed in education, leaving people behind.

> "*I use a train metaphor. So now this train is starting to move [...] Let's try to invite these people and they can catch this too. [...] If people are the first passengers, then everything will be much easier, because the later train will be very fast. You cannot catch it.*" (Participant 1)

The statement also refers to the possibility of widening pre-existing divides in QT [13] and that the fair distribution of opportunities plays a crucial role in quantum literacy [4].

**Values**

The interviews and survey reveal a clear emphasis on inclusivity, accessibility, and diversity as key values for grassroots movements in quantum. People from diverse backgrounds and experiences unanimously emphasize the importance of creating an environment that actively welcomes everyone. The results mark a contrast between the goals and values of the organizations: While one of the main goals focuses on providing educational content, their value systems don't reflect learning as a high value; instead, the values aim to democratize learning opportunities and to appeal to people from different backgrounds, aiming at inclusiveness and diversity.

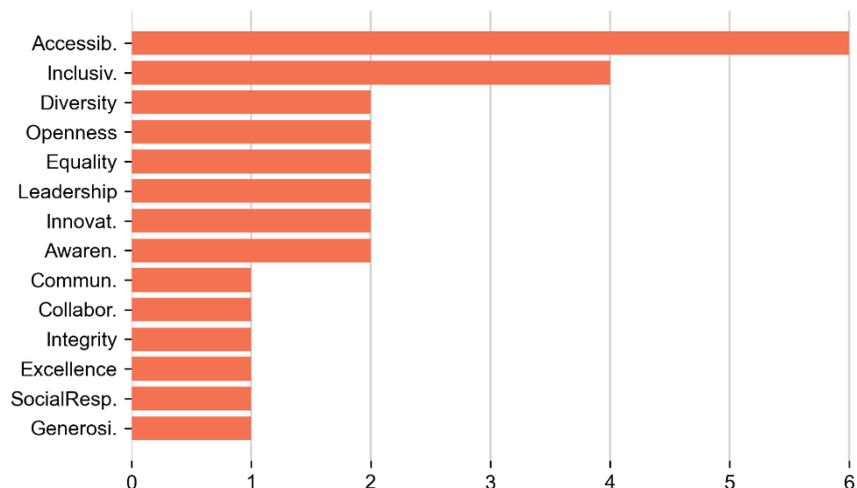

*Figure 4: ranked values, frequency count of the highest value stated.*

Of all the values mentioned, inclusiveness and accessibility were mentioned most often, followed by diversity, community, and collaboration, see Figure 5. Education and learning were mentioned 6 and 7 times, respectively. In contrast, if we look at the ranking of the highest values, learning and education are not mentioned at all, see Figure 4, instead accessibility is mentioned most often as the highest value. By prioritizing **accessibility**, grassroots organizations aim to break down barriers, ensure fair access to opportunities and learning materials, and foster a collaborative quantum community. They contribute to democratize quantum education by making quantum knowledge and resources available to everyone, regardless of background or experience. The term accessibility includes being open to all skill levels, to people with little financial resources, and being understandable, like easy explanations for starters, as well as covering many different mother tongues. Additionally, there is the accessibility to quantum computers, simulators, and research institutions that are country- or company-specific. This recognition of accessibility as a core value highlights the crucial role it plays in shaping the culture of these initiatives, making quantum research and engagement more equal and accessible to all and putting their efforts into bridging quantum divides. A quote from a global organization reflects that:

> *"We wish to foster the value of learning together through a community driven effort. And we want it all to be open source, not just in terms of like transparency, but also in terms of accessibility and inclusivity. So, we want as much about quantum computing that is accessible online to be available to people for free and available easily without barriers, be it economic barriers or other barriers, and prevent sort of restricted learning. We want to create more opportunities for people to learn the way they wish to learn about this field and eventually contribute to a skilled workforce."* (Participant 6)

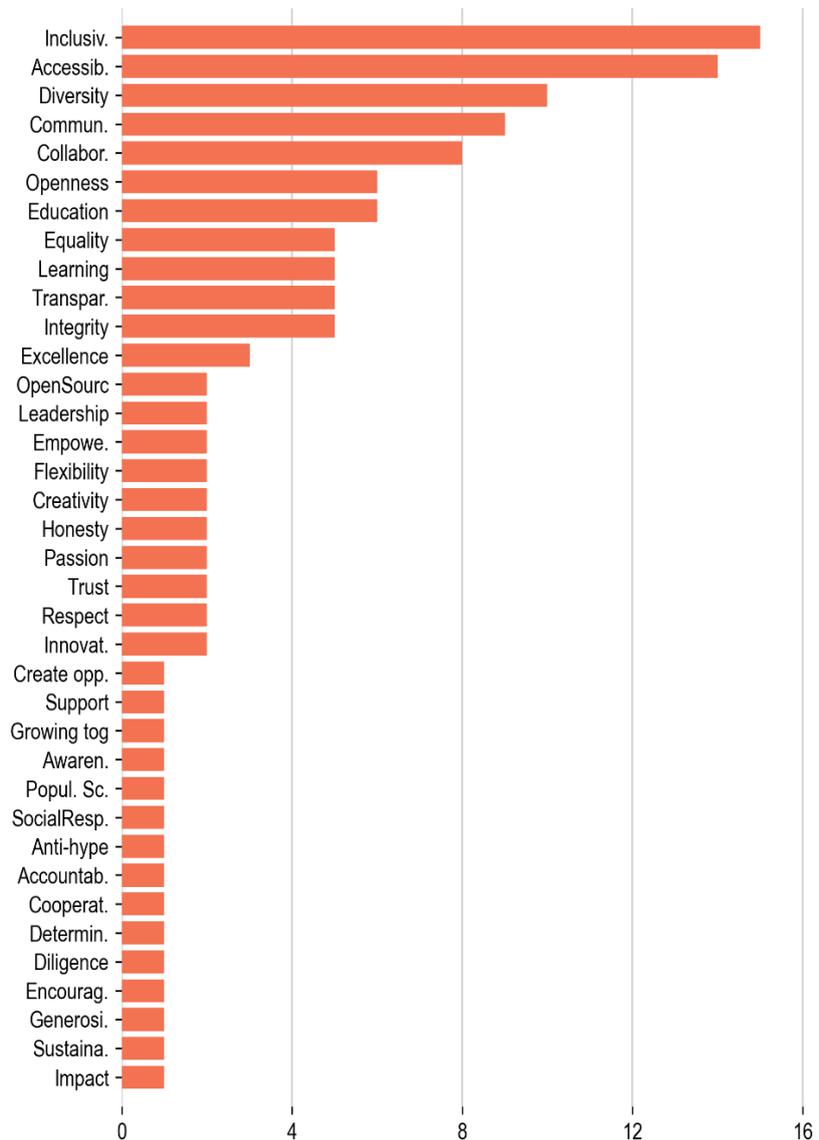

*Figure 5: Frequency count of stated values, without ranking.*

**Inclusivity and diversity**, being also positioned at the top among all mentions, are often-discussed values in the literature on quantum innovation, societal impact and workforce literacy [1, 30]. Recognizing inclusiveness and diversity as key principles underscores these organizations' commitment to leveraging different skills, cultural perspectives, and demographics for collective strength and overall benefit. The quantum grassroots talking points are best put in quotes:

*"Inclusivity is a really big value for us, because we have people from many different parts of the world who have different backgrounds, different experiences and different exposures or different levels of engagement in their communities."* (Participant 2)

*"Some values that we mainly prioritize would be, of course, one being diversity. It's kind of implied there. But we do want to see more women and minority individuals enter the playing field. That is a number one goal."* (Participant 14)

On a more structural level is this quote:

*"The reality is that the STEM workforce still struggles with the diversity and inclusivity issue. [...] we want to really make sure that we are addressing the mistakes that were made when classical computers were being developed and make sure that from the beginning of the quantum computing revolution that we're making sure that education and that the field at large is inclusive and diverse. And so, we have a big focus on making sure that we're reaching people of color, those from rural or low-income communities, women, and non-binary students, and not just make sure that we're reaching them, but that they feel empowered and included to like and feel like they belong in this space."* (Participant 15)

The research findings reveal that **community engagement** and **collaboration** emerge as important values, as they were mentioned by over half of interviewed organizations, which create (virtual) places to go to, a community, a knowledge exchange, and a network where people can find resources, answers, and a support network. The intentions behind building a community are manifold, it is about "*fostering a sense of learning together*" (Participant 6) or to "*unite with other like-minded people*" (Participant 20) and creating spaces that empower people to realize new ideas and projects. It is also a place of being motivated, inspired by others. Collaboration is not only seen as an internal value, collaborating with others, it is also about collaborating with partners, like in academia, arts, or governmental organizations and is closely related to the goal creating an ecosystem.

While **leadership** values may not be a priority for most organizations, the research identified a subset of organizations that place a high value on leadership. These organizations are driven by a collective ambition to make a meaningful contribution to the advancement of quantum science and technologies. Notably, leadership values play a crucial role in organizations operating in countries with limited quantum-specific resources, national programs, funding, educational programs, and accessibility to the latest QT.

The overall organization's value set addresses the need for a diverse quantum-literate workforce, attracting people without a STEM background and contributing to diversifying the quantum workforce overall. Moreover, this intense focus on inclusivity, accessibility, and diversity aligns with a broader societal goal of fostering innovation and progress in QT. As these organizations work to diversify the quantum workforce, they are also seeking to create a cultural shift that prioritizes fairness and equality to drive progress in quantum research and development.

**Teaching and Learning Methods**

Grassroots quantum organizations offer various learning materials in different forms to their community, most prevalently through books, videos, webinars, lectures, explanatory apps, and forums. Not only do they provide guidance and basic introductions for complete beginners, but also advanced trainings and activities for more experienced people. They offer classic lecture-style training to provide the groundwork for QT basics and theory, guided introductions, and presentations of different quantum topics, mostly related to quantum computing, to guide the learning journey and support self-learning. The depth and quality of this material can be considered as widely diverse, depending on their resources, focus or goal, strive for excellence as well as their support through collaborations and within their network. For example, an organization that focuses exclusively on providing content to learners and doing summer programs and courses have a higher dedication to creating learning resources than community-oriented organizations that use community efforts to create content and events for learners. Also, considering the average size of the quantum grassroots organizations, it strongly depends on their available capacities. But grassroots organizations are quite creative in finding different pathways: Some, rather community oriented grassroots organizations use journal clubs, offering a collection of the latest research and news articles that can be discussed within the community. They use the wisdom of their community to support academic members and PhD students who are part of the community in their respective research efforts and share knowledge openly within their organization. Since most members of global grassroots organizations are located worldwide, members in the quantum grassroots community can gain insights into quantum research from around the globe, which is mutually beneficial for academics, PhD students, and other members.

> *"We also do a journal club which is essentially taking a research paper somewhere online and breaking it down to sort of help others understand, but also encouraging them to read it up first and then sort of having a session where everyone comes in and asks questions about it and has like a roundtable discussion. […] However, we try to make it as easy as possible so that anyone can join in."* (Participant 6)

Grassroots organizations focus on collaborative learning by creating events like hackathons and learning camps to establish a network and learning environment with others and community-based forums for open discussions or curated Q&A sessions. Through events, grassroots organizations attract new members and create awareness and outreach within their ecosystem. Assignments are used to put theory into practice to check the understanding of QT concepts; hackathons focus on applied knowledge, using the quantum knowledge of the community to create problem-solving skills and developing ideas or applications of QT as well as doing community projects, applying theory to practical questions and map out solutions. Some organizations offer quantum games to learn in a playful way:

*"What we've been trying to do a lot is the educational games, […] trying to frame it in a sense of quantum rules […] and we try and take classical games that people know already and add in these quantum rules."* (Participant 5)

Most of the grassroots organizations offer a broad mix of teaching and learning styles, also reaching different learners in their needs and individual situations. They do offline and online events to connect the network more broadly, offer learning opportunities to a wider audience and keep the network connected: *"We are also trying to have more online activities. […] just a small thing where people gather. It helps us to stay in touch, stay active."* (Participant 9)

**Regional differences**

Comparing the data of the grassroots organizations from different locations around the globe, there are some differences in the value sets.

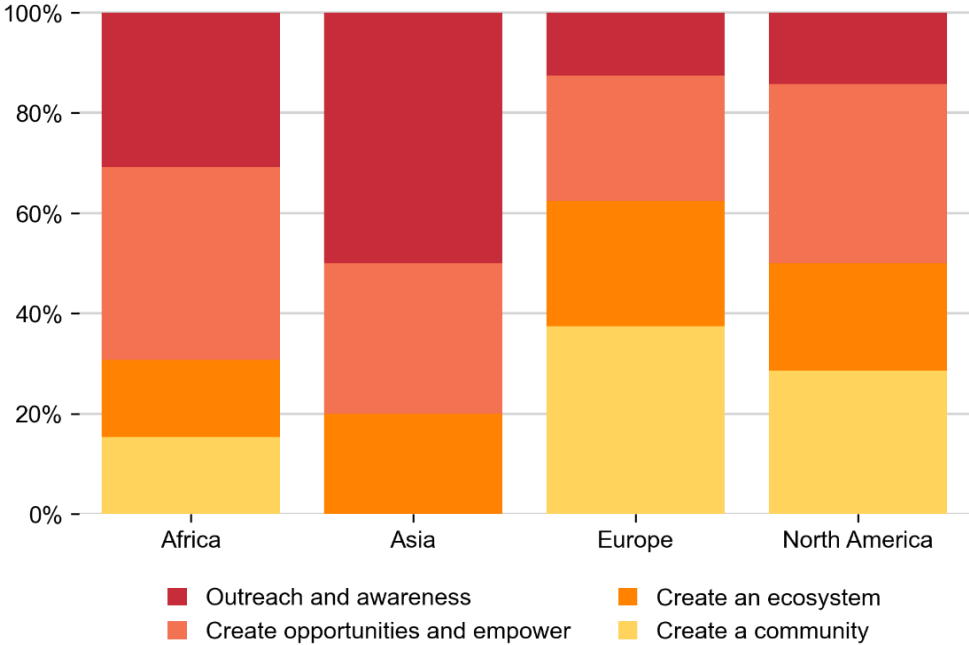

*Figure 6: Comparison of goals according to region, education was omitted as there was no difference.*

Comparing Africa and Asia with Europe and North America, there are differences in outreach and awareness. Countries with programmatic attention of QT in government, business, and education experience a stronger saturation of awareness in the ecosystems. Funding is an issue for all grassroots organizations, but the investments into QT and spending of funds are not equally distributed around the globe [9], leaving organizations from the Global South with the challenge to raise awareness and spending time educating decision-makers about QT, whereas other organizations from countries with higher investment schemes on QT can focus on community building:

*"This is the 26th year of Africa Tech Festival, and this is the first year quantum computing is a topic. So, I feel it's a big step where Africa can get to where the rest of the continent will also get to about quantum computing. I feel it creates a whole large opportunity outreach […], because there'll also be start-ups and other companies and people from government. […] I think in the next 1 to 2 years, it's something that schools could be aware of. Maybe it would still be a process of working on how to create a strong workforce in the field at continent level."* (Participant 9)

*"People will be more aware of the existence and use of quantum computing, not just as a trend, but also something that is for research and for industry."* (Participant 12)

The challenge to create awareness towards governmental or economic or even educational institutions results in explicit membership skills that quantum grassroots organizations esp. from the Global South aim for people with technical expertise and experience in government and business, as the following quote shows:

*"We try to have people on the team who are more technical to approach the more technical groups such that they speak the same language. We've also had […] people work in government already, so they know how to speak the language that people in government would be more inclined to respond to. So, then they would lead those conversations. People are good with businesses also speak to the people who are more inclined to the business side and start-ups. So, it's a team effort."* (Participant 10)

In consequence, quantum grassroots organizations from the Global South address different target groups. Most organizations address students of various ages as well as academics and professionals who want to learn, but in countries with lower investments and awareness of QT, the target groups are shifted toward decision-makers in academia, business, and governments. This strategic shift demands another degree of professionalization among the grassroots organization's members and fewer resources available to attract members and volunteers.

*"We try to organize events where we could have more Africans exposed to the technology because quantum is not yet that popular across Africa, given that it's from the 55 countries, I don't think more than 20 have a very strong quantum computing presence. So, it takes communication and networking. […] building the collaborations, we also try to get government involved. So sometimes it's trying to organize and set up those conversations with people in government, and I also try to get people in academia because they have the most access to the students who are the future, given that Africa is a large youth base."* (Participant 9)

Another reason, why countries from the Global South have a greater focus on outreach and awareness is their lack of accessibility. That is an essential driver for global grassroots organizations, providing content within the global community and bridging the gap between people who have access to exclusive learning materials, like IBM Qiskit or quantum computer simulators and those who have not.

*"Sometimes you don't have access to the information because of where you are located. But since members are in the diaspora, if they are in America, they know what some of the programs that we could benefit from are happening. So at least we have that to keep us updated on activities and events and also just the progress that's happening in the ecosystem."* (Participant 9)

Accessibility also refers to availability in different languages, showing that such barriers are especially important to global grassroots organizations and those in countries where English is less widely used. They want to reach people in their own language and introduce them to quantum content. The following quotes are from a global organization and a Brazilian one:

*"I think accessibility is definitely a big one. The fact that we have people from everywhere, a lot of our content is translated into multiple languages. It's not just English. We have Spanish or various Indian languages and sometimes we have Arabic and our channels internally. Also, we have different language channels where in the past we've had events in different languages as well. I think that speaks to accessibility is just because English may not be your first language. And for many people, Asian, it's not their first language. We don't want that to be a hindrance. We're trying to eliminate as many boundaries as possible."* (Participant 2)

*"The people speak Portuguese, there's not that many people speaking English. […] English is supposedly the universal language, but I particularly care about producing stuff in Portuguese, because my main thing about science communication is that I think there is a big gap between general public and what physicists think that is accessible."* (Participant 16)

LIMITATIONS

The study's unique focus on grassroots organizations, while valuable, does present some limitations. For instance, the selection of these organizations was confined to those operating in English, thereby excluding potential participants from Latin America and the Asian region, including China, Japan, and South Korea.

The response rate of the organizations for the survey and interviews was average at 30%, but due to the size and distribution of the member regions of the global organizations, there was a potential knowledge gap as to whether most of the organizations' members shared the value set that was mentioned by the participants. We did not find evidence in the focus groups that there is much difference, but the focus groups in this study were rather small, with only two different members of global organizations.

While regional characteristics were indeed identified based on the interview data set, it's important to note that the study's design did not specifically target regional differences. As a result, the data set is not fully representative, with significant gaps in regions such as the Middle East, China, and other countries in Asia and Latin America. However, this presents an opportunity for future research to fill these gaps and provide a more comprehensive picture.

DISCUSSION

The study's results show grassroots organizations' efforts to democratize education and its potential benefits for creating ecosystems, diversifying workforce development, and creating quantum literacy and awareness. Comparing their goals and values shows their strength in democratizing learning opportunities for people around the world, making quantum knowledge accessible, and understandable. The organizations connect private initiatives, such as IBM's Qiskit, to their members and make quantum knowledge available to students, professionals, or the local public in general. Quantum grassroots organizations use a variety of narratives to make QT accessible to a wider audience, to help build a knowledge base that is not just physics and to promote an inclusive language for the purposes of utilizing QT [5]. As Coenen et al. write "Language matters, because the way we talk about things sets a horizon of expectation for how we go about doing things." [2], which in the context of the different educational narratives and teaching methods employed of quantum knowledge offered by grassroots organizations is a valuable contribution to literacy efforts and worth exploring further. Also, the grassroots do not prioritize high quality educational materials, so their level of quality in educating students, professionals and the general public as more organizations are established remains an open question and is certainly worth exploring further.

By disseminating knowledge of QT to people of diverse backgrounds and origins, grassroots organizations play an important role in addressing the shortage of a well-trained quantum workforce [17]. Besides the fact that there are strong efforts to expand the quantum workforce through national programs, grassroots organizations contribute to addressing the quantum workforce shortage through education, by popularizing and attracting people to quantum who might not have chosen this field of interest and by creating a support network along the learning journey. By reaching out to people of all genders and backgrounds, grassroots organizations can help to diversify the future quantum workforce, benefiting societies and economies [21]. In addition, grassroots organizations build bridges while countries impose trade restrictions and focus on national agendas. Organizations from Africa or Latin America address leadership and target decision-makers to create a local quantum ecosystem that creates a need for economic and political programs, education, and funding for QT. While grassroots organizations may have little impact on changing the political and technological division between the 'haves' and 'have-nots' [13], global quantum grassroots organizations build bridges by connecting and distributing available content to countries and members with fewer technological resources and opportunities. Through their own efforts to participate in the worldwide developments of QT, they work to reduce the disparities between nations and benefit the entire quantum community [4, 14].

Grassroots efforts align closely with the values of responsible research and innovation (RRI), which supports the inclusion of a wide range of perspectives in R&D and innovation processes, as well as the consideration of ELSI of quantum. Even though impact and social responsibility were not mentioned as prior values, the organizations provide a platform and community where people with different backgrounds and needs engage in discussions about how QT could be approached and be utilized in future and anticipate different opportunities. They share their ideas in hackathons, learning events, and the creation of quantum applications, allowing for the exploration of previously inaccessible spaces of opportunity [4, 5].

The correlation between media attention to the quantum advantage and the creation of new organizations suggests that the hype around QT leads to increased public awareness and a positive impact on the number of organizations and members. However, the rise of QT performance has also attracted the interest of policymakers and entrepreneurs, leading to increased national investments and newly funded companies [8], which has led to a recalibration of expectations among scientists, emphasizing RRI and its impact on society [4, 31]. How the change of hype and narratives around QT influences grassroots organizations' values and motivation remains an open question [31, 32], as this study mostly spoke with first-generation founders and mostly long-term members. Moreover, due to the fact that a large number of new organizations were created during COVID-19, it remains an open question how the end of the pandemic will affect the community of grassroots organizations, their further establishment in the ecosystem, and the growth of new organizations.

Finally, despite the positive effects of grassroots organizations on quantum ecosystems and the future development of QT, as well as societal effects by incorporating shared values of inclusivity, diversity, and accessibility, the organizations are in the beginnings of their journey, and it is not yet clear how they develop further by managing transitions of founders and long-term members, stepping into their own careers, leaving organizational development and change to new generations. Also, the COVID-19 pandemic has passed, which has been a great accelerator for virtual meet-ups and online content [27], and openness of knowledge exchange on a global scale seems to fade. However, it is not yet clear how this affects the connectedness and strength of global and local communities, making the grassroots organizations potentially vulnerable.

CONCLUSION

In conclusion, the grassroots organizations contribute to democratizing of quantum education, making quantum knowledge accessible and understandable, to learners of all backgrounds and to countries with less resources. The numbers of founded organizations is increasing, addressing more quantum knowledge to quantum enthusiasts and attract potentially more learners worldwide. Through their work they support bridging the quantum divide, fostering quantum literacy and awareness also among the public and enable participation and engagement of diverse perspectives in the quantum ecosystem and future QT development.

Grassroots organizations address the shortage of a quantum literate workforce, by offering a broad mix of educational resources and teaching methods, providing platforms for discussions and explorations as well as their strong focus on diversity and inclusion. They bridge current gaps for non-STEM students as well as learners from different backgrounds, approaching minorities, inviting them into the quantum realm, widening the potential future workforce. Their efforts on diversity and inclusion are beneficial to the quantum ecosystem, but also to societies in general, increasing a broad knowledge base, which potentially results in higher trust-building, responsible innovation efforts, ethical, legal, and social considerations and even more diverse quantum applications.

Despite the positive impacts, these organizations face challenges, such as managing transitions among founders and long-term members and adapting to post-pandemic shifts in global connectedness and knowledge exchange. Also, while the hype around QT has spurred media attention, public awareness, and national investments, it remains to be seen how the changing environment towards market formation and rivalry, will influence grassroots organizations' values and motivations. As these organizations continue to evolve, their ability to sustain and grow with their core values will determine their future impact on the global quantum ecosystem.

ACKNOWLEDGMENT

The initial work for this study was conducted as part of the QIntern2023 program at QWorld with the support of volunteer interns. We would like to thank them for their assistance throughout the course of this study.